\newif\ifisonecolumn
\isonecolumntrue %use this to make it true (or comment if not)

\ifisonecolumn
\documentclass[onecolumn,draft,12pt]{IEEEtran}
\else
\documentclass{IEEEtran}
\fi

\input{defi_tran_ULDL.def}

\usepackage{amssymb}

\usepackage{etoolbox}
\newcommand{\FigDat}[2]{
\ifstrequal{#1}{System}{
\begin{figure}[t]
        \begin{center}\includegraphics[scale=#2]{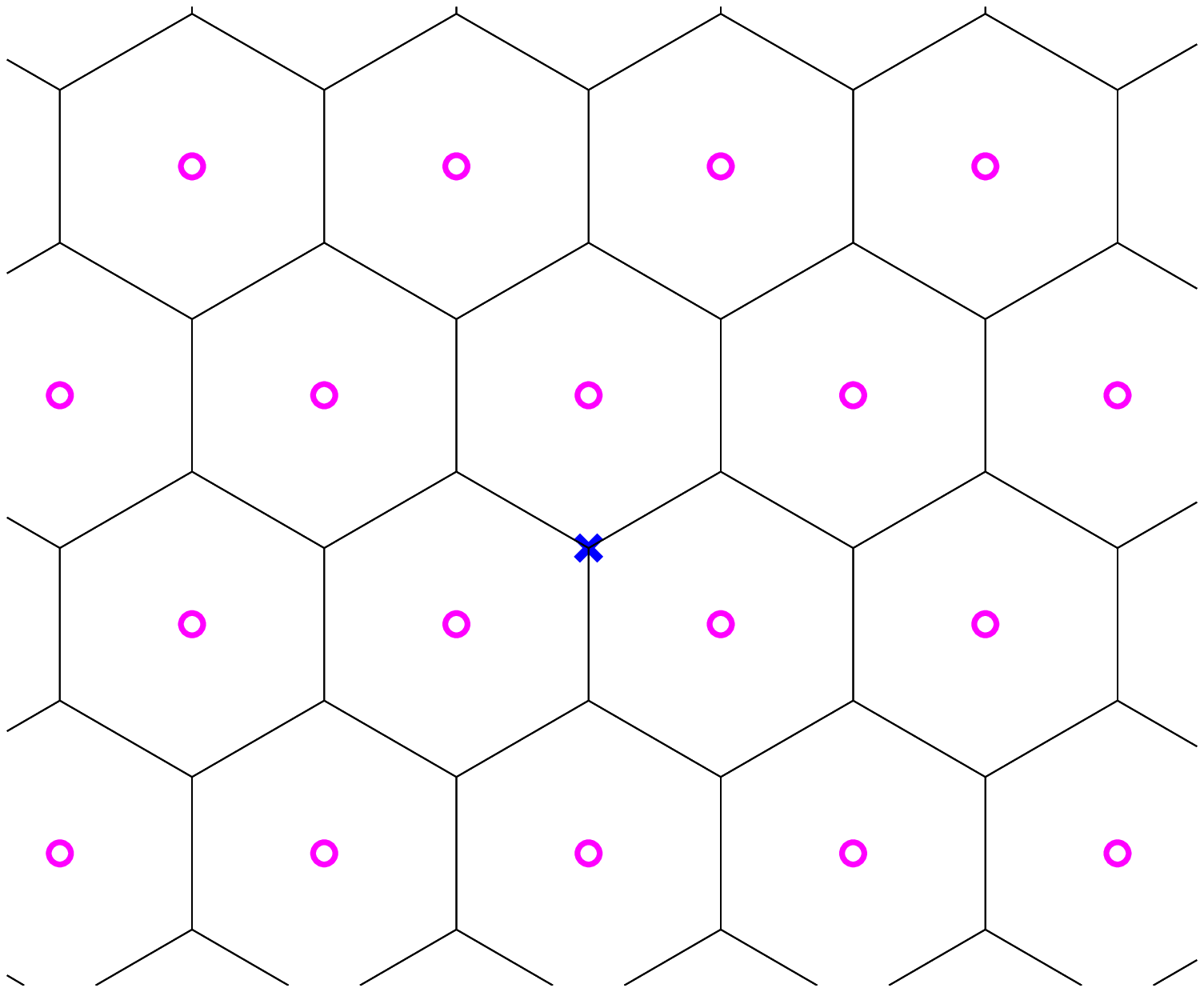}\end{center}
    \caption{Cellular system model. BSs are denoted by circles, the tested mobile is denoted by an x-mark, and the cell borders are marked by solid lines.}
    \label{f:system model}
\end{figure}
}{}
\ifstrequal{#1}{Performance}{
\begin{figure}[t]
        \begin{center}\includegraphics[scale=#2]{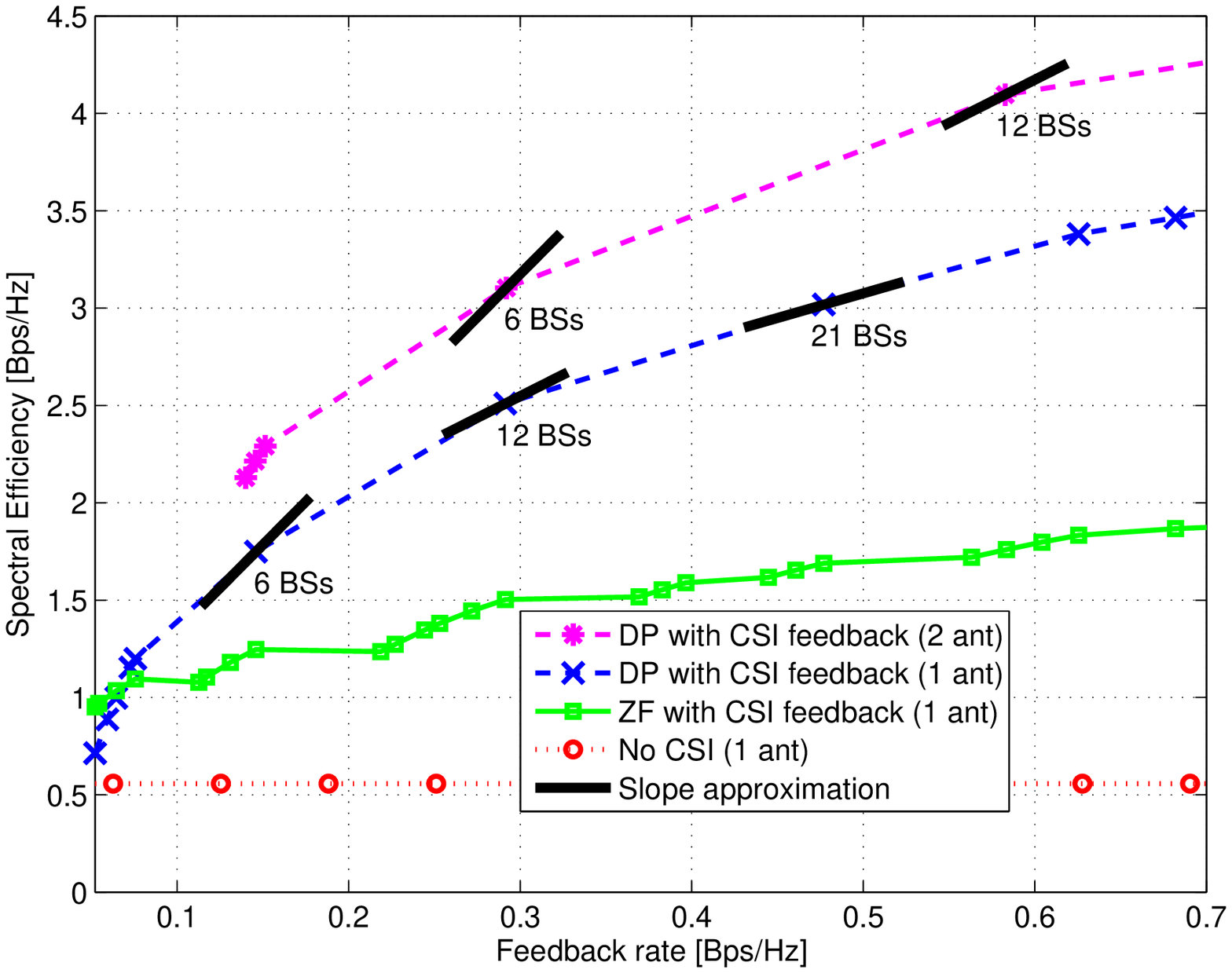}\end{center}
    \caption{Average spectral efficiency per user in the downlink of the cellular network, as a function of the spectral efficiency used for feedback in the uplink.}
    \label{f:balance fig}
\end{figure}
}{}
\ifstrequal{#1}{mult_BS_select}{
\begin{figure}[t]
        \begin{center}\includegraphics[scale=#2]{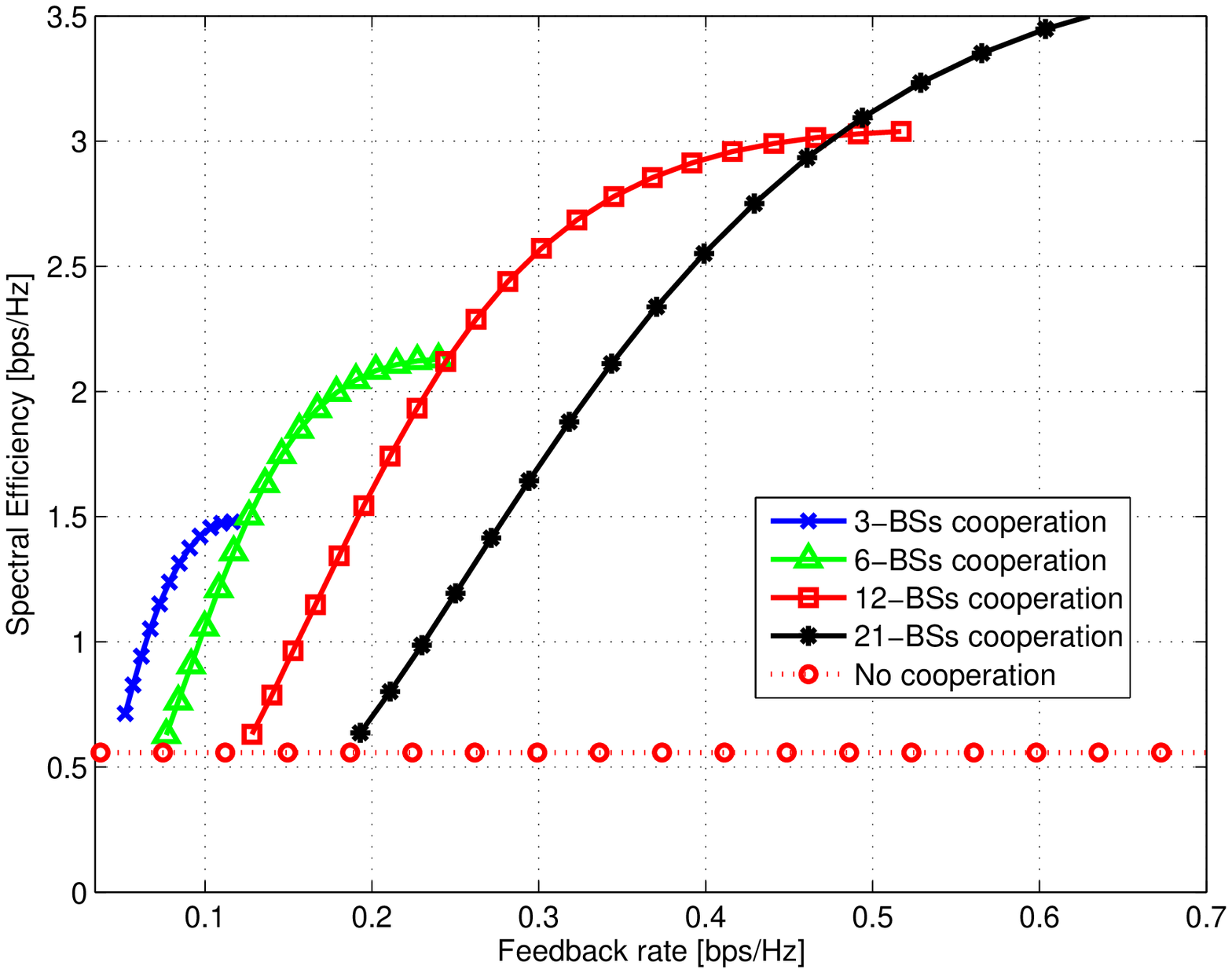}\end{center}
    \caption{Average spectral efficiency per user in the downlink of the cellular network for various cooperation scenarios, as a function of the spectral efficiency used for feedback in the uplink.}
    \label{f:multch fig}
\end{figure}
}{}
\ifstrequal{#1}{VsRho}{
\begin{figure}[t]
        \begin{center}\includegraphics[scale=#2]{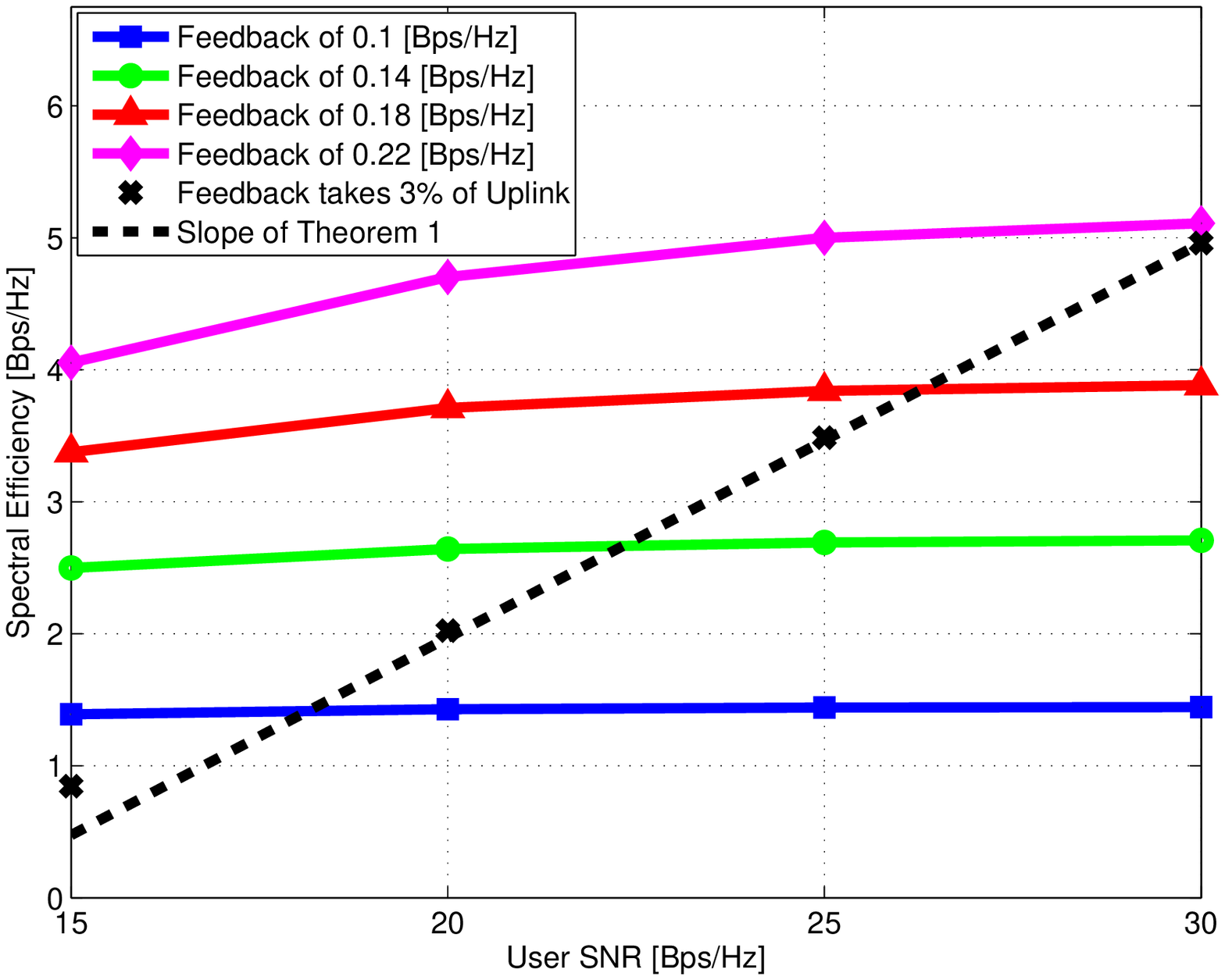}\end{center}
    \caption{Average spectral efficiency per user in the downlink of the cellular network as a function of the SNR per user from the nearest BS. The figure shows the performance for fixed feedback rates and for adaptive rate.}
    \label{f:VsRho}
\end{figure}
}{}
}
\begin{document}

\title{Uplink Downlink Rate Balancing in Cooperating Cellular Networks}
\author{Itsik~Bergel,~\IEEEmembership{Senior Member,~IEEE,}
        Yona~Perets,
        and~Shlomo~Shamai,~\IEEEmembership{Fellow,~IEEE}% <-this % stops a space
\thanks{\copyright {} 2015 IEEE. Personal use of this material is permitted. Permission from IEEE must be obtained for all other uses, in any current or future media, including reprinting/republishing this material for advertising or promotional purposes, creating new collective works, for resale or redistribution to servers or lists, or reuse of any copyrighted component of this work in other works.}% <-this % stops a space
\thanks{I. Bergel is with Faculty of Engineering, Bar-Ilan University, 52900 Ramat-Gan, Israel;
   (e-mail: bergeli@biu.ac.il).}% <-this % stops a space
\thanks{Y. Perets is with Marvell Semiconductor Israel, Azorim Park, Petach-Tikva, 49527, Israel; (e-mail: yoni@marvell.com).}% <-this % stops %a space
\thanks{S. Shamai (Shitz) is with the Department of Electrical Engineering, Technion-
Israel Institute of Technology, Technion City, Haifa 32000, Israel; (e-mail:
sshlomo@ee.technion.ac.il)}% <-this % stops a space
\thanks{Manuscript submitted December, 2014; Revised May, 2015.}}

\maketitle

\begin{abstract}
Broadcast MIMO techniques can significantly increase the throughput in the downlink of cellular networks, at the price of channel state information (CSI) feedback from the mobiles, sent over the uplink.
Thus, these techniques create a mechanism that can tradeoff some uplink capacity for increased downlink capacity. In this work we quantify this tradeoff and study the exchange ratio between the feedback rate (over the uplink) and the downlink rate. We study both finite and infinite networks, and show that for high enough (but finite) SNR, the uplink rate can be exchanged for increased downlink rate with a favorable exchange ratio. This exchange ratio is an increasing function of the channel coherence time, and a decreasing function of the number of measured base stations.
We also show that in finite networks, devoting a constant fraction of the uplink to CSI feedback can increase the downlink multiplexing gain continuously from $0$ to $1$, as a function of the fraction size. On the other hand, in infinite networks (with infinite connectivity) our lower bound is only able to show doubly logarithmic scaling of the rate with SNR.
The presented results prove that the adaptation of the feedback rate can control the balance between the uplink and downlink rates. This capability is very important in  modern cellular networks, where the operators need to respond to continuously changing user demands.
\end{abstract}

%%%%%%%%%%%%%%%%%%%%%%%%%%%%%%%%%%%%%%%%%%%%
\section{Introduction}

The limited availability of spectrum resources on one hand, and the exponential growth of wireless data traffic on the other hand generate a strong motivation to improve the spectral
efficiency per unit area offered by wireless systems. As an example, the overall data traffic over cellular networks grew by more than an order of magnitude during just the past four years
\cite{ericsson2013Nov,NorthAmerica2011LTE}, and this trend is expected to continue and intensify.

The conventional approach of deploying more base stations per unit area is
not sufficient to satisfy the huge growth in traffic demand.
Therefore, more advanced techniques such as cooperative transmission and interference mitigation at the transmitter side (also known as broadcast MIMO or multi-user MIMO techniques)  are being considered
(e.g., \cite{ramprashad2011cooperative}) and standardized (e.g., \cite{gpp36819}) for the next generation of cellular networks.

Broadcast MIMO techniques can significantly increase the capacity when the transmitter has more antennas than each of the receivers (e.g., \cite{vishwanath2003duality}\nocite{spencer2004introduction}--\cite{sharif2005capacity}). The term Broadcast MIMO typically applies to the case of a single transmitter, but, it is applicable also to multi cell transmission if the inter-cell cooperation is good enough (see for example \cite{simeone2011cooperative} and references therein).

Some interesting recent works have shown capacity gain using delayed feedback \cite{maddah2012completely}, or even with no feedback, if the network can ensure effective channel variations at predetermined times \cite{gou2011aiming}. Yet the more significant (and more thoroughly researched) capacity gain of broadcast MIMO techniques depends on the availability of (at least partial) timely  channel state information (CSI). In this work we focus on the downlink of FDD cellular networks, where partial CSI can be obtained by feedback from the mobiles (over the uplink).

The capacity gain in multi-user MIMO can be divided into two types. The first, commonly termed \textit{multi-user diversity}, results from the selection of good active users out of a large number of users (e.g., \cite{yoo2007multi}--\nocite{zhang2007mimo,shirani2010mimo}\cite{ khoshnevis2011limited}). In this work we focus on the second type of capacity gain, which results from the spatial multiplexing of several data streams to several different users. Note that Ravindran and Jindal \cite{ravindran2012transactions} have shown that for a limited feedback rate this second type (i.e., sending good CSI feedback for a subset of users) is more efficient than multi-user diversity.

Due to the importance of feedback in these types of schemes, the problem of feedback design and optimization has been studied extensively (e.g., 
\cite{love2004value}\nocite{gesbert2007shifting,love2008overview,heath2008exploiting,
chae2008coordinated,guthy2008finite,trivellato2008transceiver,kobayashi2011training}--\cite{kakishima2012evaluating}).
Jindal \cite{jindal2006mimo} has shown that downlink capacity depends on the feedback rate, and that in order to achieve full multiplexing gain, the feedback rate must scale as the logarithm of the SNR. Further works on this subject have affirmed this scaling relation and presented various feedback schemes to better exploit the channel (e.g. \cite{kountouris2006multiuser}--\nocite{ravindran2008limited,caire2010multiuser,ayach2012interference}\cite{wagner2012large}).

However, all analysis of broadcast MIMO channels with limited feedback have so far focused only on the downlink, and did not attempt to quantify the tradeoff between uplink and downlink rates.

In cellular communication systems there is a constant need to balance the uplink and downlink rates according to the user demands. In most cases the required downlink rates are much higher than the required uplink rates (e.g., \cite{falaki2010first,ericsson2013})\footnote{Typically, the uplink and downlink capacities of a FDD network are also different, and a typical downlink capacity can be up to twice the UL capacity. Yet this difference often does not compensate for the huge traffic asymmetry, which can even be as high as a factor of 10.}. In such cases the operator would prefer to tradeoff some of the uplink rate in order to increase the downlink rate. Furthermore, the ratio between the demand for uplink rate and downlink rate constantly changes. Thus, there is an acute need for a capability to control the uplink-downlink rate balance.

In time division duplex (TDD) systems, such a tradeoff can be established through the control of the ratio between the time devoted to uplink and the time devoted to downlink. In frequency division duplex (FDD) systems this tradeoff can be achieved by controlling the bandwidths allocated for uplink and downlink. But, in order not to create interference, the changes (in both cases) must be identical over the whole network, which limits the possibility to dynamically balance the uplink and downlink rates. Furthermore, in FDD,  changing the bandwidth allocation is often impossible due to regulation constraints and equipment capabilities. 

The application of broadcast MIMO techniques creates the ability to tradeoff between a rate increase in the downlink and a rate loss in the uplink (as part of it is devoted to feedback). In this work we present a convenient method to balance the uplink and downlink rates using a feedback mechanism in a broadcast MIMO network.

The performance of the balancing mechanism is characterized by quantifying the tradeoff between the downlink rate and the uplink rate. We show that by using broadcast MIMO techniques we can control the uplink-downlink ratio by a proper adaptation of the feedback rate.
The main concept of this work was originally described in \cite{yellin2011balancing,bergel2012IEEEI}. For finite networks in the interference limited regime, we show that uplink rate can be exchanged for downlink rate with a constant exchange ratio, which depends on the size of the network and on the channel coherence time. We also show that devoting a constant fraction of the uplink to CSI feedback can vary the downlink multiplexing gain continuously from $0$ to $1$ (as a function of the fraction size).

In infinite networks the situation is more complicated. In order to increase the downlink throughput, each mobile is required to send feedback on  the channels of an increasing number of base stations (BSs). This increase in the number of BSs changes the asymptotic behavior of the downlink rate, and our lower bound is only able to show a logarithmic scaling of the downlink rate with the feedback rate (and not linear with the rate).
Yet for practical SNRs, each mobile only sends feedback on the channels from a fairly small number of BSs, and the rate loss compared to the finite network is not large. This loss is quantified in Section \ref{sec: uplink downlink tradeoff} by an approximation of the ratio between  the slope of the downlink rate with respect to the feedback rate in an infinite network and the same slope in the finite network that sends feedback on the same number of channels.

It is often assumed that the linear zero-forcing (ZF) scheme enables one to achieve close to the MIMO broadcast channel capacity. In mathematical terms, it was shown that the high SNR expression of the sum rate in the ZF scheme differs from the channel capacity only by a constant term that is independent of the SNR \cite{lee2007high}. It is important to note, however, that the constant difference of \cite{lee2007high} does depend on the network size. As a result, we show in this work that for infinite networks the performance of the linear ZF scheme is quite poor. Hence, the main focus of this work is on the (non-linear) dirty paper scheme \cite{costa1983writing}.

This work is based on a simple and practical approach, which uses a robust lower bound on the achievable rates. In this way, we establish a useful scheme for uplink-downlink rate balancing, which can be adapted
according to user demands and network operator priorities.

The remainder of this paper is organized as follows: In Section \ref{sec: system model} we present the system model and constraints. In Section \ref{sec: uplink downlink tradeoff} we present and discuss our main results on the
tradeoff between uplink and downlink rates. The proof of the results is given in Section \ref{sec: proof}, which includes the details of the analyzed transmission scheme and channel quantization schemes. Section \ref{sec: numeric} presents a numerical
example and Section \ref{sec: discussion} contains our concluding remarks.

%%%%%%%%%%%%%%%%%%%%%%%%%%%%%%%%%%%%%%%%%%%%
\section{System model \& Preliminaries}\label{sec: system model}
We consider the cellular network with multiple BSs having a total of $M$ antennas, which serve $N\le M$ single antenna mobiles.  Assuming a flat fading channel and focusing on the downlink,  the symbols that are received by all the mobiles at a given time can be written in a vector form as:
\begin{IEEEeqnarray}{rCl}\label{d: channel model}
\yv = \sqrt{\rho}\Hm\xv +\nv
\end{IEEEeqnarray}
where $\Hm$ is the $N \times M$ complex valued channel matrix, whose $i,j$-th element, $\ch_{i,j}$,  is the channel gain from the $j$-th BS antenna to the $i$-th mobile, $\xv$ is an $M \times 1$ complex vector of
the transmitted symbols from all BSs, and $\nv$ is an $N \times 1$ noise vector whose elements are mutually independent proper complex Gaussian random variables \cite{neeser1993proper}
with zero mean and unit variance. We assume that the transmitted symbols satisfy:
\begin{IEEEeqnarray}{rCl}\label{e: x power constraint}
E[\|\xv\|^2] \le N 
\end{IEEEeqnarray}
and hence $\rho$ can be considered as the normalized transmission power (i.e., transmission power divided by noise power). 

In the following we will mostly focus on an analysis of the achievable rates as the normalized transmission power grows. This type of analysis represents the growth in received SNR that was observed in recent years due to the deployment of more BSs and the decrease in the average distance between BSs and mobiles. Thus, we assume that the normalized transmission power of the uplink and the downlink grow at the same rate, with their relationship described by:
\begin{IEEEeqnarray}{rCl}\label{e:k_UL}
\rho_{\rm UL}=\kappa_{\rm UL}\cdot \rho
\end{IEEEeqnarray}
where $\kappa_{\rm UL}$ captures the ratio between the transmission power of the mobile to the transmission power of the BS, as well as the different noise figures at the receive chains.

We assume a block fading channel model, so that $\Hm$ is assumed to be fixed during the transmission of a block of $T$ symbols, but changes in an independent manner from block to block. If the channel is frequency selective, we assume that some type of multi-carrier modulation is adopted (for example, orthogonal frequency
division multiplexing, OFDM). In such a case, we assume that the channel is fixed during $T_\textrm{t}$ OFDM symbols and also identical for a block of $B$ frequency bins. We thus have a coherence block of $T=T_\textrm{t}\cdot B$ symbols.

The channel matrix $\Hm$ represents the effects of path loss, long term fading and short term fading. The channel elements are assumed to be mutually independent, proper, complex Gaussian random variables with zero mean, giving rise to the very popular Rayleigh fading channel model that is often used to model the short-term fading of wireless links (e.g., \cite{sklar1997rayleigh}). The variance of the  $i,j$-th channel element is denoted by
\begin{IEEEeqnarray}{rCl}\label{e: channel fading power}
E[|\ch _{i,j}|^2] = \sigma^{2}_{i,j} .
\end{IEEEeqnarray}
The values of $\sigma^2_{i,j}$ are constant throughout the analysis, and represent the topology related path-loss effects and the long-term fading effects (e.g., log-normal shadowing).

Our focus is on full BS cooperation (i.e., zero-delay, infinite capacity backhaul links between all the BSs in the system), perfect receiver CSI at the mobiles\footnote{Perfect receiver CSI is an oversimplification when the number of
estimated channels becomes very large (e.g., \cite{jose2011pilot}--\nocite{hoydis2011massive,yin2012coordinated}\cite{lozano2013fundamental}). The effect of imperfect receiver CSI is not considered in this work, and is left for future study.}. On the other hand, the BSs only have quantized CSI, which limits their ability to coordinate their transmissions and avoid interference between the data of different mobiles. Due to the availability of zero-delay, infinite capacity backhaul links, the network performance  depend only on the total number of antennas and their channel gains, while the distribution of the $M$ antennas between the BSs is irrelevant. Thus, we will address only the total number of BS's antennas, and will not consider the actual number of BSs.

In order to assess the load of the quantized CSI feedback on the up-link rate, we shall make use of the following lower bound on the achievable {\em uplink} rate per mobile user in a Rayleigh fading cellular network
(obtained for example from the results of \cite{vishwanath2003duality,goldsmith2003capacity}):
\begin{IEEEeqnarray}{rCl}\label{e: Rul lower bound}
R_{\rm UL}\ge\frac{1}{N}E\left[\log_2\det\left(I+\rho_{\rm UL}  \Hm_{\rm UL}^H\Hm_{\rm UL}\right)\right].
\end{IEEEeqnarray}
where the $M \times N$ matrix $\Hm_{\rm UL}$ describes the uplink channels between the $N$  mobiles and the $M$  antennas, $I$ denotes the identity matrix and $\rho_{UL}$ is the normalized transmission power for each of the mobiles.
As in the downlink case, we shall assume the common Rayleigh fading model for the uplink channel, with different gains to different channel elements so as to take into account path-loss and long-term fading effects as before.

%%%%%%%%%%%%%%%%%%%%%%%%%%%%%%%%%%%%%%%%%%%%
\section{Uplink-Downlink rate tradeoff}\label{sec: uplink downlink tradeoff}
Our main results characterize the relation between the feedback rate and the achievable downlink rate. These results, which are summarized by Theorem \ref{Th: Fixed feedback rate finite} and Theorem \ref{Th: Fixed feedback rate infinite}, are derived by optimizing the allocation of the allowed feedback rate to the various quantized channels. The theorems consider two different scenarios. In the finite network scenario, for high enough feedback rate, it turns out to be best to send feedback on the channel gains of \textbf{all} antennas in the network. Thus, the main question is how much feedback to allocate to the quantization of the channel of each antenna. On the other hand, in the infinite network scenario, the mobile cannot send feedback on the channels from all antennas. Hence, we face an additional question: Which channels should be quantized and transmitted on the uplink to optimally balance between the residual interference of the un-quantized antennas and the quantization errors of the quantized antennas.

Define by $F_i$ the total feedback rate sent by mobile $i$ (over the uplink) and by $R_i$ the downlink rate achievable by mobile $i$. 
\begin{theorem}[Limited connectivity scenario]\label{Th: Fixed feedback rate finite} If mobile $i$ only receives signals from a finite set of antennas, $\sL_i$, of cardinality $L_i$ (i.e., $\sigma^{2}_{i,j}=0 \forall j\notin\sL_i$) then  the ratio between the downlink rate  and the corresponding feedback rate in the interference limited regime is lower bounded by:
\begin{IEEEeqnarray}{rCl}\label{e: th1_bound}
\lim\limits_{\rho \rightarrow \infty}\frac{R_i}{F_i}\ge\frac{T}{L_i}.
\end{IEEEeqnarray}
Furthermore, if the feedback rate is set to occupy a constant fraction of the uplink channel, i.e., $F_i=r \cdot R_{\rm UL}$ then:
\begin{IEEEeqnarray}{rCl}\label{e: Th2 bound}
\lim_{\rho \rightarrow\infty}\frac{R_i}{\log_2 \rho }\ge\min\left\{1,\frac{r \cdot T}{L_i} \right\}.
\end{IEEEeqnarray}
\end{theorem}
\begin{theorem}[Infinite network with infinite connectivity]\label{Th: Fixed feedback rate infinite} In an infinite network ($N\rightarrow \infty$) with path loss exponent of $\alpha > 2$ and no shadowing (i.e., $\sigma^{2}_{i,j}=\rho \cdot d_{i,j}^{-\alpha}$ where $d_{i,j}$ is the distance between the $i$-th mobile and the $j$-th antenna) the asymptotic ratio between the downlink rate  and the logarithm of the corresponding feedback rate  is lower bounded by:
\begin{IEEEeqnarray}{rCl}\label{e: Th3 bound}
\lim_{F_i\rightarrow \infty}\lim\limits_{\rho \rightarrow \infty}\frac{R_i}{\log_2F_i}\ge\frac{\alpha}{2}-1.
\end{IEEEeqnarray}
and if the feedback rate is set to occupy a constant fraction of the uplink channel, i.e., $F_i=r \cdot R_{\rm UL}$ then:
\begin{IEEEeqnarray}{rCl}\label{e: Th4 bound}
\lim_{\rho \rightarrow\infty}\frac{R_i}{\log_2\log_2 \rho }\ge \frac{\alpha}{2}-1.
\end{IEEEeqnarray}
\end{theorem}
\begin{IEEEproof}[Proof of theorems] See Section \ref{sec: proof}.
\end{IEEEproof}
\

The bounds of Theorems \ref{Th: Fixed feedback rate finite} and \ref{Th: Fixed feedback rate infinite} obviously hold for the case of optimal quantization. However, to make the result more robust, we also proved that these bounds hold for the simplest case of scalar uniform quantization. This point is further emphasized in Section \ref{sec: numeric}, where we demonstrate that the presented bounds characterize the achievable performance in practical communication systems.

One should note that the results of both theorems are only lower bounds on the achievable performance. These lower bounds help us to prove that the rate balancing principle is indeed feasible. The presented bounds represent the best scaling currently  known for the MIMO broadcast channel. Yet, at the current time there is no (non-trivial) upper bound for the MIMO broadcast channel  in the presented scenario. Hence, the optimal rate scaling remains an open question.

The results in Theorem \ref{Th: Fixed feedback rate finite} are very appealing. As stated above,  FDD cellular networks have an acute need for an uplink-downlink rate balancing mechanism. Equation (\ref{e: th1_bound}) shows that there is a possibility to tradeoff uplink rate (feedback) to downlink rate at a constant `price'. Furthermore, in many practical scenarios this `price' is quite low as the channel coherence time and coherence bandwidth are large.

Consider for example the ETSI pedestrian B channel \cite{etsi1998101} at a pedestrian speed of $3$Km/H, and using a carrier frequency of $2$GHz. The pedestrian B  channel is characterized by a delay spread of $t_d=630$nS and Doppler spread of $f_d\approx 5.5$Hz. Thus, its coherence time is proportional to $T_\textrm{t}\sim 1/f_d$ and its coherence bandwidth is proportional to $B\sim 1/t_d$.
However, the translation between the pedestrian B channel model (or any practical channel model) and the analyzed block fading model is not straightforward. The pedestrian B model considers continuous variation of the channel both in time and in frequency, while our model consider blocks in time  and frequency in which the channel does not change at all. 

To relate the two models, we need to define a sensitivity threshold, and choose the coherence time and coherence bandwidth such that the channel will not change more than this threshold within a coherence block (see \cite{bergel2012bounds} for a deeper discussion on the relation between the coherence models).
In the following, we choose to be conservative and limit the channel variations in a coherence block to $-20$dB. Considering the autocorrelation function of the channel impulse response, a proportionality constant  of $1/40$ is required to satisfy the $-20$dB threshold. Thus,  the coherence block for the feedback evaluation is $T=T_\textrm{t}\cdot B=\frac{1}{40 f_d}\cdot \frac{1}{40  t_d} \approx 180$ symbols. 
 
 The `price' for increasing the spectral efficiency of the downlink by $1$ Bps/Hz is $L_i/T$ Bps/Hz of feedback over the uplink. Due to the size of $T$, this is a reasonable cost in practical systems even if the mobile will send feedback to $L_i=100$ antennas.

The results of Theorem \ref{Th: Fixed feedback rate infinite} may seem more pessimistic, as it shows only a logarithmic increase of the downlink rate with the feedback rate in an infinite network (with infinite connectivity). Moreover, the scaling of the rate with SNR is no longer logarithmic, but only doubly logarithmic. While the results give only a lower bound and the upper bound is not yet known, it is important to note that this is the first result that shows that an infinite network with infinite connectivity and partial CSI is not interference limited. In other words, the rate increase with SNR is not bounded although each receiver is interfered by an infinite number of antennas. Furthermore, these results should not be discouraging because the doubly-logarithmic behavior only  characterizes the behavior at extremely high SNRs.

To demonstrate the behavior of the infinite network in finite (but high) SNR, we derive in Appendix \ref{app: proof of corrolary on rate slope} the following approximation for the derivative of the bound: 
\begin{proposition}\label{cor: cor 1}In the setting of the second part of Theorem \ref{Th: Fixed feedback rate infinite}, there exists a transmission scheme that uses a feedback rate $F_i\le \tilde F_i(\rho) \le r \cdot R_{\rm UL}$, sends feedback to $\tilde L_i(\rho)$ antennas, and achieves a rate  $R_i\ge\tilde R_i(\rho)$ such that: $ \tilde R_i(\rho)$ achieves the bound of Theorem 2, i.e., 
\begin{IEEEeqnarray}{rCl}
\lim_{\rho \rightarrow\infty}\frac{\tilde R_i(\rho)}{\log_2\log_2 \rho }\ge \frac{\alpha}{2}-1.
\end{IEEEeqnarray}
and also
\begin{IEEEeqnarray}{rCl}\label{e: inifinite network slope}
\lim_{\rho\rightarrow\infty}\frac{  \tilde L_i(\rho) }{T }\frac{d\tilde R_i(\rho)}{d \tilde F_i}&=&\left(1-\frac{2}{\alpha}\right)\cdot\frac{\alpha \log_2\left(e \right)}{\alpha \log_2(e)+2Q}
\end{IEEEeqnarray}
with $Q=0$ for optimal quantization and $Q=2$ for scalar quantization.
\end{proposition}
\begin{IEEEproof}[Proof of Proposition \ref{cor: cor 1}] See Appendix \ref{app: proof of corrolary on rate slope}.
\end{IEEEproof}

Proposition \ref{cor: cor 1} addresses the same setup as Theorem \ref{Th: Fixed feedback rate infinite} and allows us to better understand the tradeoff between uplink rate and downlink rate. Using Proposition \ref{cor: cor 1} we can say that (for high enough feedback rate) the gain in downlink rate from an increase of $1$bps/Hz in feedback rate can be approximated by the slope:
\begin{IEEEeqnarray}{rCl}\label{e: inifinite network slope apx}
\frac{d\tilde R_i}{d\tilde F_i}&\simeq&\left(1-\frac{2}{\alpha}\right)\frac{T }{  \tilde L_i(\tilde F_i) }\cdot\frac{\alpha \log_2\left(e \right)}{\alpha \log_2(e)+2Q}.
\end{IEEEeqnarray}
This result looks quite similar to the result of Theorem \ref{Th: Fixed feedback rate finite}, and hence seems more optimistic than Theorem \ref{Th: Fixed feedback rate infinite}.  

Comparing (\ref{e: inifinite network slope apx}) and the result for the finite network model of (\ref{e: th1_bound}) shows that the derivative in the two cases differs only by a constant factor. Hence, the main measure for the efficiency of the feedback is the number of antennas for which the mobile sends feedback, $\tilde L_i(\tilde F_i)$, compared to the coherence block size, $T$. As the coherence block, $T$, is typically very large, the linear tradeoff results of (\ref{e: th1_bound}) and (\ref{e: inifinite network slope apx}) hold even if the mobile sends feedback for hundreds of antennas, which is way beyond the capabilities of any state of the art network.

The proof of Theorems \ref{Th: Fixed feedback rate finite} and \ref{Th: Fixed feedback rate infinite} provides the mathematical justification for the claims above. Moreover, this proof is a constructive proof that is based on a robust and appealing transmission scheme - dirty paper coding using quantized channel matrices, where quantization errors are treated as additional noise. The proof is presented in the following section.

%%%%%%%%%%%%%%%%%%%%%%%%%%%%%%%%%%%%
\section{Proof of Theorems \ref{Th: Fixed feedback rate finite} and \ref{Th: Fixed feedback rate infinite}}\label{sec: proof}
The proof is based on an optimization of the feedback rate in a network with cooperative interference mitigation at the BSs. We start with the description of the feedback scheme and the transmission scheme, and then present the feedback rate optimization and evaluate the achievable downlink rate.

\subsection{Channel quantization scheme}
As mentioned above, we assume that each mobile has  perfect knowledge of the channel. Each mobile quantizes its measured channels and sends the index of the quantized version to the BSs. We write the channel matrix as the sum:
\begin{IEEEeqnarray}{rCl}\label{d: quantization error representation}
\Hm =\hHm +\eM
\end{IEEEeqnarray}
where $\hHm$ is the quantized channel known to the transmitter, while $\eM$ is the channel error that is not known to the transmitter. In the following we detail two possible quantization schemes. In both schemes the matrices    $\hHm$ and  $\eM$ are statistically independent and also all elements of $\eM$ are zero mean and statistically independent. Denoting the $i,j$-th element of $\eM$ by $\eMe_{i,j}$ and its variance by $\XMe_{i,j}^2$ we have:
\begin{IEEEeqnarray}{rCl}\label{e: e uncorrelated}
E\left[\eMe_{i_1,j_1}\eMe_{i_2,j_2}^*\right]=
\begin{cases}
\XMe_{i,j}^2 & i_1=i_2\textrm{ and }j_1=j_2
\\
0 & \textrm{otherwise}
\end{cases}.
\end{IEEEeqnarray}
Combining with (\ref{e: channel fading power}), we also know that:
\begin{IEEEeqnarray}{rCl}
E\left[|\hat \ch _{i,j}|^2\right] = \sigma^{2}_{i,j}-\XMe_{i,j}^2.
\end{IEEEeqnarray}

In the following we will consider two extreme quantization schemes. Define by $F_{i,j}$ the total rate allocated for the feedback on the channel of antenna $j$ from mobile $i$, we will show that for both schemes the feedback rate is given by:
\begin{IEEEeqnarray}{rCl}\label{e: feedback rate formula}
F_{i,j}\le\begin{cases}   \frac{1}{T}\left[Q+\log_2 \left( \frac{\sigma_{i,j}^2 }{\XMe_{i,j}^2}\right)\right] & \XMe^2 _{i,j}<\sigma_{i,j}^2
\\ 0 & \XMe^2 _{i,j}=\sigma_{i,j}^2
\end{cases}
\end{IEEEeqnarray}
with $Q=0$ for optimal quantization and $Q=2$ for scalar quantization.
\subsubsection{Optimal quantization}
 Using rate distortion theory, for statistically independent, proper complex Gaussian channel coefficients,
the rate required to quantize a sample of variance $\sigma_x^2$ to a quantization mean square error of $D$ is \cite{berger1971rate,cover2006elements}:
\begin{IEEEeqnarray}{rCl}
R(D)=\begin{cases}
\log_2\left(\frac{\sigma_x^2}{D}\right) & D\le\sigma_x^2 \\
0 & D>\sigma_x^2
\end{cases}.
\end{IEEEeqnarray}
Thus, we have:
\begin{IEEEeqnarray}{rCl}\label{e: optimal feedback rate formula}
F_{i,j}=\begin{cases}\frac{1}{T}
\log_2\left(\frac{\sigma_{i,j}^2 }{\XMe^2 _{i,j}}\right) & \XMe^2 _{i,j}<\sigma_{i,j}^2
\\ 0 & \XMe^2 _{i,j}=\sigma_{i,j}^2
\end{cases}
\end{IEEEeqnarray}
where $\XMe_{i,j}^2=E[|\epsilon_{i,j}|^2]$ is the quantization error variance, and the division by $T$ is due to the fact that the feedback is assumed to be sent once for each $T$ consecutive channel uses (the coherence block).
Obviously, we have $\XMe_{i,j}^2\le\sigma^{2}_{i,j}$, and the case where $\XMe^{2}_{i,j}=\sigma^{2}_{i,j}$ corresponds to the scenario in which mobile $i$ does not send feedback on antenna $j$.
Furthermore, by the asymptotic properties of the rate distortion function (e.g., \cite{cover2006elements} chapter 10.3), each quantization error element, $\epsilon_{i,j}$, is a proper complex Gaussian variable with zero mean, which is statistically independent of the quantized channel, $\hHm$, the transmitted symbols, $\xv$
and every other element of $\eM$.

\subsubsection{Scalar quantization}\label{subsub: scalar quantization}
As a simple alternative, we consider also a scalar uniform quantizer. This quantizer has a fixed step size of $\Delta_{i,j}$ and it quantizes the real and imaginary parts of the channel gain to the nearest multiple of $\Delta_{i,j}$.  In order to guarantee the statistical independence between the quantization values and the quantization error, we add to the channel gain a dither signal, $d$, which is uniformly distributed over the square $-\Delta_{i,j}/2<\Re(d)\le\Delta_{i,j}/2$ and $-\Delta_{i,j}/2<\Im(d)\le\Delta_{i,j}/2$ (where $\Re(d)$ and $\Im(d)$ represent the real and imaginary parts of the complex variable $d$, respectively). This dithering also guarantees that the quantization error will have a uniform distribution, and hence the quantization error is given by: $\XMe_{i,j}^2=\Delta_{i,j}^2/6$.

To evaluate the amount of feedback required, we use a lower bound on the entropy of the feedback. Gish and Pierce \cite{gish1968asymptotically} showed that asymptotically, as the quantization step approaches zero, the entropy of a scalarly quantized (real) Gaussian random variable is $0.255+\frac{1}{2}\log_2 \left( \frac{\sigma^2 }{\XMe^2}\right)$. Thus, if we limit the discussion to a limited range of quantization errors, we can upper bound the resulting entropy of a scalarly quantized complex Gaussian random variable by $Q+\log_2 \left( \frac{\sigma^2 }{\XMe^2}\right)$, using an appropriately chosen constant $Q$. Noting that there is no point in quantizing the channel with a quantization error that is larger than the channel power (i.e., when $\XMe_{i,j}^2\ge\sigma_{i,j}^2  $), in the following we need an upper bound for any $\XMe_{i,j}^2<\sigma_{i,j}^2 $. For the dithered scalar quantization described above and the range $\XMe_{i,j}^2<\sigma_{i,j}^2 $, one can easily verify that the desired upper bound is obtained using $Q=2$, i.e., for a complex Gaussian channel gain $H_{i,j}$ with variance $\sigma^{2}_{i,j}$, the required feedback rate is upper bounded by:
\begin{IEEEeqnarray}{rCl}\label{e: scalar feedback rate formula}
F_{i,j}\le\begin{cases}   \frac{1}{T}\left[2+\log_2 \left( \frac{\sigma_{i,j}^2 }{\XMe_{i,j}^2}\right)\right] & \XMe^2 _{i,j}<\sigma_{i,j}^2
\\ 0 & \XMe^2 _{i,j}=\sigma_{i,j}^2
\end{cases}
\end{IEEEeqnarray}
Note that this feedback bound has a constant gap of $2$ bits from the optimal quantization feedback rate. 

\subsection{Downlink transmission scheme}

The lower bound presented here is based  on the dirty paper encoding scheme with zero forcing linear pre-processing (DP-ZF). In this scheme the transmitter tries to completely remove the interference at the price of signal power loss  \cite{ginis2002vectored}.
This scheme is known to be close to optimal at high SINR \cite{caire2003achievable}.
The bound also assumes that all antennas transmit at the same power, $\rho $. Such homogenous power distribution was shown to be suboptimal in the high SNR regime, in the case of partial BS cooperation \cite{bergel2012linear}. It is also sub-optimal in the case of complete BS cooperation if the feedback rate does not increase with SNR. On the other hand, for high enough feedback rate, homogenous power distribution is close to optimal in the high SNR regime.

Following the DP-ZF scheme, we use of the LQ decomposition of the quantized channel matrix:
\begin{IEEEeqnarray}{rCl}\label{e: LQ decomp}
\hHm=\Lm \Qm
\end{IEEEeqnarray}
where $\Lm$ is a $N\times M$ lower triangular matrix and $\Qm$ is a $M\times M$ unitary matrix.
We apply a precoding of the data symbols:
\begin{IEEEeqnarray}{rCl}
\xv=\Qm_N^H \sv
\end{IEEEeqnarray}
where $\Qm_N$ is the $N\times M$ matrix that contain the top $N$ rows of $\Qm$,  $\xv$ is the vector of transmitted symbols and $\sv$ contains the actual data symbols. The resulting effective channel is:
\begin{IEEEeqnarray}{rCl}\label{e: revised channel model}
\yv =  \sqrt{\rho}\Lm \sv + \sqrt{\rho}\eM\xv +\nv.
\end{IEEEeqnarray}
Recalling that $\Lm$ and $\sv$ are known to the BSs while  $\eM$ is not known,
we define:
\begin{IEEEeqnarray}{rCl}
z_i= \sqrt{\rho}\sum_{j=1}^{i-1} \ell_{i,j}s_j
\end{IEEEeqnarray}
where $ \ell  _{i,j}$ is the $i,j$-th element of the matrix $\Lm$, and
\begin{IEEEeqnarray}{rCl}\label{e: effective noise term}
w_i=n_i+ \sqrt{\rho}\sum_{j=1}^M \eMe_{i,j}x_j .
\end{IEEEeqnarray}
Thus, we can write:
\begin{IEEEeqnarray}{rCl}\label{e: Lemma channel model}
y_i =  \sqrt{\rho}\ell_{i,i} s_i +z_i+w_i .
\end{IEEEeqnarray}

The zero mean of $\eMe_{i,j}$ and its statistical independence on $\hHm$ guarantee that $E[s_i w_i]=E[z _i w_i]=0$. Hence, 
we can use the results of \cite{Bergel2014Letter,Bergel2014SPAWC}, which showed that considering the unknown interference as additional white noise results in a lower bound on the achievable rate. Thus, the channel capacity is lower bounded by:
\begin{IEEEeqnarray}{rCl}\label{e: lemma 1 eq}
R
&\ge&
E\left[
\log_2\left(1+\frac{ \rho|\ell_{i,i} |^2 E[|s_i|^2]}{E[|w_i|^2]}\right)
\right] .
\end{IEEEeqnarray}
Note that the proof of \cite{Bergel2014SPAWC} is a constructive proof, i.e., it presents a scheme that can actually achieve the  lower bound.

 Setting $E[\sv\sv^H]= I$, which complies with the power constraint, and evaluating the variance of $w_i$ in (\ref{e: effective noise term}) results in:
\begin{IEEEeqnarray}{rCl}\label{e: Ri from full version}
R_i
&\ge& E\left[\log_2\left(1+\frac{\rho | \ell  _{i,i}|^2}{1+\rho \cdot\sum_{j=1}^M \XMe_{i,j}^2}\right)\right].
\end{IEEEeqnarray}

Note that Equation (\ref{e: Ri from full version}) extends the DP-ZF scheme \cite{ginis2002vectored} to the case of partial CSI, at the price of additional interference to each receiver. Thus, Equation (\ref{e: Ri from full version}) provides a convenient lower bound on the achievable data rates which is particularly important in the case of small channel errors. In the following we use this bound to
 quantify the uplink-downlink rate balancing tradeoff.

\subsection{Quantization optimization}
We next consider the rate allocation problem of the $i$-th mobile, i.e., the optimal division of the allowed feedback rate, $F_i$,  to the feedback of the quantization values of the channels from the various antennas.
 As seen from (\ref{e: Ri from full version}), for a given total rate of $F_i$, the objective is to minimize $\sum_j\XMe^2 _{i,j}$. Using (\ref{e: feedback rate formula}) the optimization problem can be written as:
\begin{IEEEeqnarray}{rCl}
\min_{\sum_j F_{i,j} \le F_i} \sum_j \sigma_{i,j}^2  2^{ U(F_{i,j})\cdot (Q-F_{i,j}T)}.
\end{IEEEeqnarray}
where $U(x)=1$ if $x>0$ and $0$ otherwise.
Using the Lagrange multiplier, $\lambda$, we have:
\begin{IEEEeqnarray}{rCl}
\Lambda_i(\lambda)=\sum_j \sigma_{i,j}^2  2^{U(F_{i,j})\cdot (Q-F_{i,j}T)}+\lambda\sum_j F_{i,j}.
\end{IEEEeqnarray}
Due to the discontinuity at $F_{i,j}=0$, we need to distinguish between active and non active feedback links. According to (\ref{e: feedback rate formula}) using $F_{i,j}T<Q$ will results in a worse error than without any feedback. Thus, the set of antennas for which mobile $i$ sends feedback is given by:
\begin{IEEEeqnarray}{rCl}\label{e: sL(F_i) definition}
\sL_i(F_i)=\{j: F_{i,j}T>Q\}.
\end{IEEEeqnarray}For any $j\in \sL_i(F_i)$, taking the derivative with respect to  $F_{i,j}$:
\begin{IEEEeqnarray}{rCl}
\frac{\partial \Lambda_i(\lambda)}{\partial F_{i,j}}=-\frac{T\sigma_{i,j}^2 }{\log_2(e)}\cdot 2^{Q-F_{i,j}T}+\lambda.
\end{IEEEeqnarray}
and equating to zero, we get the optimal feedback rate:
\begin{IEEEeqnarray}{rCl}\label{e: optimal feedback rate}
F_{i,j}T=\log_2 \sigma_{i,j}^2 -\lambda'
\end{IEEEeqnarray}
where $\lambda'$ is a constant that is chosen to satisfy the sum feedback constraint.
In the case that $Q=0$, $\lambda'$ also determines the activity on the feedback link, and $j\in \sL_i(F_i)$ if $\log_2 \sigma_{i,j}^2 -\lambda'>0$. In this case, the optimal rate allocation is very much like the water pouring algorithm used for power allocation in parallel Gaussian channels \cite{gallager1968information}. Thus, $\lambda'$ can represent a water level and the feedback is allocated according to the gap between $\log_2 \sigma_{i,j}^2$ and the water level. Antennas for which $\log_2 \sigma_{i,j}^2$ is below the water level will not get any feedback from mobile $i$. 

On the other hand, if $Q>0$ there is an additional cost for the activation of a feedback link. Hence, the activity is determined by a different parameter, $\lambda_\textrm{A}$. Thus, in general 
\begin{IEEEeqnarray}{rCl}\label{e: set cal L definition}
j\in \sL_i(F_i)$ if $ \sigma_{i,j}^2 >\sigma_0^2 \triangleq 2^{\lambda_\textrm{A}}
\end{IEEEeqnarray}

 and $\lambda_\textrm{A}\ge\lambda'$.   
  
Denote by $L_i(F_i)=|\sL_i(F_i)|$ the size of the set of the active feedback links. We have:
\begin{IEEEeqnarray}{rCl}
F_i T=\sum_{j\in\sL_i(F_i)}\log_2 \sigma_{i,j}^2 - L_i(F_i) \cdot\lambda'
\end{IEEEeqnarray}
and the Lagrange multiplier can be evaluated by:
\begin{IEEEeqnarray}{rCl}\label{e: lambda' value}
\lambda'=\frac{\sum_{j\in\sL_i(F_i)}\log_2 \sigma_{i,j}^2 -F_i T}{ L_i(F_i) }.
\end{IEEEeqnarray}
Substituting in (\ref{e: optimal feedback rate}) we have for any $j \in \sL_i(F_i)$:
\begin{IEEEeqnarray}{rCl}\label{e: F_ij T}
F_{i,j}T=\log_2 \sigma_{i,j}^2 -\frac{\sum_{j\in\sL_i(F_i)}\log_2 \sigma_{i,j}^2 -F_i T}{ L_i(F_i) }
\end{IEEEeqnarray}
and using (\ref{e: feedback rate formula}):
\begin{IEEEeqnarray}{rCl}\label{e: equal quantization error}
\XMe^2 _{i,j}&=&\sigma_{i,j}^2  \cdot2^{\left(Q+\frac{\sum_{j\in\sL_i(F_i)}\log_2 \sigma_{i,j}^2 -F_i T}{ L_i(F_i) }-\log_2 \sigma_{i,j}^2 \right)}
\nonumber\\
 &=&2^{\left(Q+\frac{\sum_{j\in\sL_i(F_i)}\log_2 \sigma_{i,j}^2 -F_i T}{ L_i(F_i) }\right)}=2^{Q+\lambda'}=\xi^2_i(F_i)\IEEEeqnarraynumspace
\end{IEEEeqnarray}
for any $j\in\sL_i(F_i)$.
Note that (\ref{e: equal quantization error}) shows that the quantization error power is independent of the antenna's index, $j$, i.e., the quantization error power is identical for all quantized antennas.

\subsection{Rate lower bound}
\subsubsection{Proof of Theorem \ref{Th: Fixed feedback rate finite} - Limited connectivity scenario}
In this scenario, mobile $i$ is only affected by a finite set of antennas, $\sL_i$, with size $L_i=|\sL_i|$, i.e., $\sigma_{i,j}^2=0$ for each $j\notin \sL_i$. As shown above, the size of the set of the active feedback links, $L_i(F_i)$ increases as the allocated feedback rate, $F_i$, increases\footnote{This is more intuitive for $Q=0$, where $\lambda_\textrm{A}=\lambda'$ and the relation between the set size and the feedback rate is established by (\ref{e: set cal L definition}) and (\ref{e: lambda' value}). The case of $Q>0$ is more difficult to visualize. Yet, the feedback set will include any affecting antenna if the feedback rate is large enough.}. Thus, for large enough feedback rate, the mobile will send feedback to all connected antennas ($\lim_{F_i \rightarrow \infty}L_i(F_i)= L_i$). Thus, there exists a rate, $\underline{F}$, so that $L_i(F_i)=L_i$ for every $F_i>\uF$.
The sum of the quantization errors becomes:
\begin{IEEEeqnarray}{rCl}
\sum_j \XMe_{i,j}^2=L_i\cdot\xi^2_i(F_i)=L_i\cdot2^{\left(Q+\frac{\sum_{j}\log_2 \sigma_{i,j}^2 -F_i T}{ L_i }\right)}\end{IEEEeqnarray}
and
the downlink rates, (\ref{e: Ri from full version}), are lower bounded by:
\begin{IEEEeqnarray}{rCl}\label{e: Ri lower bound with quantization}
R_i \TwoColHspace{-0.5mm}
\ge\TwoColHspace{-0.7mm} E\left[\log_2\Bigg(\TwoColHspace{-0.7mm}1\TwoColHspace{-0.3mm}+\TwoColHspace{-0.3mm}\frac{\rho | \ell  _{i,i}|^2}{1+\rho  L_i\cdot 2^Q\cdot 2^{\frac{\sum_j\log_2 \sigma_{i,j}^2 }{L_i}}\cdot 2^{-\frac{F_i T}{L_i}}}\Bigg)\right]\TwoColHspace{-1mm}.\IEEEeqnarraynumspace
\end{IEEEeqnarray}
Inspecting the denominator of (\ref{e: Ri lower bound with quantization}) one can derive the asymptotic result of Jindal \cite{jindal2006mimo}, which showed that in order to achieve the full precoding gain, the feedback rate should increase as $F_i \sim \log_2(\rho )$.
In this work we do not aim to achieve the full precoding gain. Instead we are interested in controlling the balance between the uplink and downlink channels.

Defining the limit of (\ref{e: Ri lower bound with quantization}) as $\rho $ goes to infinity as $\breve R_i(F_i)$ we have $\lim_{\rho \rightarrow \infty} R_i\ge \breve R_i(F_i)$ with:
\begin{IEEEeqnarray}{rCl}\label{e: first p limit}
\breve R_i(F_i)= E\left[\log_2\Bigg(\TwoColHspace{-0.7mm}1\TwoColHspace{-0.3mm}+\TwoColHspace{-0.3mm}\frac{ | \ell  _{i,i}|^2}{  L_i\cdot 2^Q\cdot 2^{\frac{\sum_j\log_2 \sigma_{i,j}^2 }{L_i}}\cdot 2^{-\frac{F_i T}{L_i}}}\Bigg)\right]\TwoColHspace{-1mm}.\IEEEeqnarraynumspace
\end{IEEEeqnarray}
Dividing by $F_i$ and taking the limit as $F_i$ goes to infinity can easily show that (\ref{e: th1_bound}) holds for high enough feedback rates. To prove (\ref{e: th1_bound}) for any feedback rate, we next apply a time sharing argument. Assume that mobile $i$ achieves an average feedback rate of $F_i$ by using a rate of $F_i/\eta$ for a fraction, $\eta$, of the time, and sending no feedback the rest of the time. Thus, mobile $i$ can achieve:
\begin{IEEEeqnarray}{rCl}\label{e: time sharing feedback}
\lim_{\rho \rightarrow \infty} \frac{R_i}{F_i}&\ge& \max_\eta\frac{(1-\eta)\breve R_i(0)+\eta \breve R_i(F_i/\eta)}{F_i}
\nonumber \\
&\ge& \lim_{\eta\rightarrow 0}\frac{ \breve R_i(F_i/\eta)}{F_i/\eta}
= \lim_{F_i \rightarrow \infty}\frac{ \breve R_i(F_i)}{F_i}.
\end{IEEEeqnarray}
Substituting (\ref{e: first p limit}) into (\ref{e: time sharing feedback}) and taking the limit as $F_i$ goes to infinity completes the proof of (\ref{e: th1_bound}) for any feedback rate.

Next, we consider a scheme in which the feedback rate is proportional to the uplink rate, using $F_i=r \cdot R_{\rm UL}$. From (\ref{e: Rul lower bound}) we have:
\begin{IEEEeqnarray}{rCl}\label{e: Fi lower bound for fixed ratio}
F_i&\ge&\frac{r}{N}E\left[\log_2\det\left(I+\rho \kappa_{\rm UL}  \Hm_{\rm UL}^H\Hm_{\rm UL}\right)\right]
\TwoOneColumnAlternate{\nonumber \\ &=&}{=}r\log_2 \rho  +g(\rho )\IEEEeqnarraynumspace
\end{IEEEeqnarray}
where $g(\rho )$ satisfies:
\begin{IEEEeqnarray}{rCl}
\lim_{\rho \rightarrow \infty} g(\rho )=\frac{r}{N}E\left[\log_2\det\left(  \kappa_{\rm UL}  \Hm_{\rm UL}^H\Hm_{\rm UL}\right)\right].
\end{IEEEeqnarray}
Substituting (\ref{e: Fi lower bound for fixed ratio}) in (\ref{e: Ri lower bound with quantization}) we have:
\begin{IEEEeqnarray}{rCl}\label{e: Ri with g(p)}
R_i&\ge&
E\Bigg[\log_2 \Bigg(
\TwoOneColumnAlternate{\nonumber \\ &&}{}
1+\frac{\rho | \ell  _{i,i}|^2}{1+\rho  L_i\cdot 2^Q\cdot 2^{\frac{\sum_j\log_2 \sigma_{i,j}^2 }{L_i}}\cdot \rho ^{-\frac{r T}{L_i}}2^{-\frac{T \cdot g(\rho )}{L_i}}}\Bigg)
\Bigg].\IEEEeqnarraynumspace
\end{IEEEeqnarray}
Unlike the limit in (\ref{e: first p limit}), in this case the limit as $\rho $ goes to infinity differs in the noise limited case ($r T/L_i\ge1$) where the denominator converges and the interference limited case ($r T/L_i<1$) where the denominator is not bounded. Thus, the limit of (\ref{e: Ri with g(p)}) is given by:
\begin{IEEEeqnarray}{rCl}
\lim_{\rho \rightarrow\infty}\frac{R_i}{\log_2 \rho }\begin{cases}=1 & r\ge\frac{L_i}{T}\\\ge\frac{r T}{L_i} & r<\frac{L_i}{T}\end{cases}.
\end{IEEEeqnarray}
which completes the proof of (\ref{e: Th2 bound}) and Theorem \ref{Th: Fixed feedback rate finite}.

\subsubsection{Proof of Theorem \ref{Th: Fixed feedback rate infinite} - Infinite network with infinite connectivity}\label{subsub: proof inifinite}
The case of an infinite network with infinite connectivity between each antenna to each mobile is slightly more complicated, due to the unbounded growth of the number of quantized antennas as the feedback rate grows to infinity. 

The maximization of the network capacity given a total feedback rate requires an optimization with respect to the parameters $\sigma_0^2$ and $\xi_i^2$. As we derive a lower bound, we simplify the derivation by choosing a sub-optimal setup in which $\sigma_0^2=\xi_i^2$. This choice is optimal for the optimal quantization case ($Q=0$), but will cause some capacity loss in the scalar quantization case ($Q>0$). 
Note that for this choice, both the data rate, $R_i$ and feedback rate, $F_i$ are monotonic decreasing functions of $\xi_i^2$ (although not necessarily continuous functions).
Thus, instead of directly maximizing the downlink rates, $R_i$, for given feedback rates, $F_i$, we characterize the network by the mobiles quantization errors, $\xi_i^2$, and study the resulting downlink rates and feedback rates.

Using this choice of the optimization parameters, the definition of the set $\sL_i(F_i)$, (\ref{e: set cal L definition}), can be written as:
$\sL_i(F_i)=\{j:\sigma^2_{i,j}>\xi_i^2\}$. Using also  the channel exponential decay law of Theorem \ref{Th: Fixed feedback rate infinite} gives:
\begin{IEEEeqnarray}{rCl}\label{e: modified set definition}
\sL_i(F_i) &=&\{j:d_{i,j}<\xi_i^{-2/\alpha}\}.
\end{IEEEeqnarray}

To further characterize the set size, we note that for any planar network with finite density and for each mobile, $i$,  there exists a constant $b_i$ so that
 the number of antennas in a circle of radius $d$ around mobile $i$  is upper bounded by:
\begin{IEEEeqnarray}{rCl}\label{e: circle bound}
N(d)\le b_i d^2.
\end{IEEEeqnarray}
Substituting in (\ref{e: modified set definition}), the set size is upper bounded:
\begin{IEEEeqnarray}{rCl}\label{e: Li upper bound}
L_i(F_i)\le \lfloor b_i \xi_i^{-4/\alpha} \rfloor.
\end{IEEEeqnarray}

Without loss of generality we will assume hereon that the antennas are sorted by their distance to mobile $i$. Thus, the distance of antenna $j$ from mobile $i$ is lower bounded by:
\begin{IEEEeqnarray}{rCl}
d_{i,j}\ge \sqrt{\frac{j}{b_i}}
\end{IEEEeqnarray}
and the channel power from this antenna is bounded by:
\begin{IEEEeqnarray}{rCl}\label{e: radius power bound}
\sigma^{2}_{i,j}\le \left(\frac{j}{b_i}\right)^{-\alpha/2}.
\end{IEEEeqnarray}
In this case, the
set of quantized antennas is necessarily given by: $\sL_i(F_i)=\{1,2,\ldots,L_i(F_i)\}$, and the feedback rate from mobile $i$ can be upper bounded:
\begin{IEEEeqnarray}{rCl}
F_i&=&\sum_{j=1}^{\infty} F_{i,j}\le \frac{Q \cdot L_i(F_i)}{T}+\frac{1}{T}\sum_{j=1}^{L_i(F_i)}
\log_2\left(\frac{\sigma_{i,j}^2 }{\xi^2_{i}}\right)
\nonumber \\
&\le&\frac{Q \cdot   \lfloor b_i  \xi_i^{-4/\alpha} \rfloor}{T}+\frac{1}{T}\sum_{j=1}^{\lfloor b_i \xi_i^{-4/\alpha} \rfloor}\log_2\left(\frac{ j^{-\alpha/2}}{b_i^{-\alpha/2} \xi^2_{i}}\right)
\nonumber \\
&=&\frac{\alpha}{2T}\sum_{j=2}^{\lfloor b_i \xi_i^{-4/\alpha} \rfloor}\left(-\log_2\left(j\right)\right)
\TwoOneColumnAlternate{\nonumber \\ &&}{}
+\frac{\lfloor b_i \xi_i^{-4/\alpha} \rfloor}{T}\left(Q+\frac{\alpha}{2}\log_2\left(b_i \xi_i^{-4/\alpha}\right)\right).
\end{IEEEeqnarray}
Noting that the term in the sum is a decreasing function of $j$, it can be upper bounded by the integral:
\begin{IEEEeqnarray}{rCl}\label{e: F_i upper bound infinite network (start)}
F_i&\le&\frac{\alpha}{2T}\int_{1}^{\lfloor b_i \xi_i^{-4/\alpha} \rfloor}\left(-\log_2\left(j\right)\right)dj
\TwoOneColumnAlternate{\nonumber \\ &&}{}
+\frac{\lfloor b_i \xi_i^{-4/\alpha} \rfloor}{T}\cdot\frac{\alpha}{2}\log_2\left(b_i 2^{2Q/\alpha}\xi_i^{-4/\alpha}\right)
\nonumber \\
&=&-\frac{\alpha }{2T}    \lfloor b_i \xi_i^{-4/\alpha} \rfloor \left(\log_2(\lfloor b_i \xi_i^{-4/\alpha} \rfloor)-\log_2(e)\right)
\nonumber \\&&
-\frac{\alpha }{2T}\log_2(e)+\frac{\alpha \lfloor b_i \xi_i^{-4/\alpha} \rfloor}{2T}\log_2\left(b_i 2^{2Q/\alpha}\xi_i^{-4/\alpha}\right).\IEEEeqnarraynumspace
\end{IEEEeqnarray}
Using the inequality $\log x \le x-1$ for $x \ge 1$, we have $y(\log \frac{x}{y}-\frac{x}{y}+1)\le 0$ for any $0\le y \le x$. This can be stated as:
\begin{IEEEeqnarray}{rCl}
-y (\log_2 y -\log_2 e) \le x \log_2 e -y\log_2 x.
\end{IEEEeqnarray}
Substituting in (\ref{e: F_i upper bound infinite network (start)}) using $x= b_i \xi_i^{-4/\alpha}$ and $y=\lfloor x \rfloor$ 
results in the upper bound
\begin{IEEEeqnarray}{rCl}\label{e: F_i upper bound infinite network}
F_i&\le&\frac{\alpha \log_2(e)}{2T}\left(  b_i \xi_i^{-4/\alpha}  -1\right)+\frac{  \lfloor b_i \xi_i^{-4/\alpha} \rfloor Q }{T}
\nonumber \\
&\le&\frac{\alpha \log_2(e)}{2T}\left(  b_i \xi_i^{-4/\alpha}  -1\right)+\frac{ b_i \xi_i^{-4/\alpha}Q }{T}.
\end{IEEEeqnarray}

We also need an upper bound on the sum of the residual interference:
\begin{IEEEeqnarray}{rCl}\label{e: sum XMe first upper bound}
\sum_{j=1}^{\infty}\XMe^2 _{i,j}&=&L_i \xi^2_i + \sum_{j=L_i+1}^{\infty} \sigma^2_{i,j}
\nonumber \\
&\le&\lfloor b_i \xi_i^{-4/\alpha}\rfloor\xi^2_i+\sum_{j=\lfloor b_i \xi_i^{-4/\alpha}\rfloor+1}^{\infty} \left(\frac{j}{b_i}\right)^{-\alpha/2}
\end{IEEEeqnarray}
where we used (\ref{e: Li upper bound}), (\ref{e: radius power bound}), and the fact that $\sigma_{i,j}^2\le\xi_i^2$ for any $j>L_i$. Using again an integral to upper bound the sum in (\ref{e: sum XMe first upper bound}), for $b_i \xi_i^{-4/\alpha}\ge 1$:
\begin{IEEEeqnarray}{rCl}\label{e: sum XMe second (ibtegral) bound}
\sum_{j=1}^{\infty}\XMe^2 _{i,j}&\le&
\lfloor b_i \xi_i^{-4/\alpha}\rfloor\xi^2_i+\int_{j=\lfloor b_i \xi_i^{-4/\alpha}\rfloor}^{\infty} \left(\frac{j}{b_i}\right)^{-\alpha/2} dj
\nonumber \\
&=&
\lfloor b_i \xi_i^{-4/\alpha}\rfloor\xi^2_i-\frac{b_i^{\alpha/2}}{1-\frac{\alpha}{2}}\lfloor b_i \xi_i^{-4/\alpha}\rfloor ^{1-\alpha/2}
\nonumber \\
&=&
\lfloor b_i \xi_i^{-4/\alpha}\rfloor\xi^2_i\cdot\left(1-\frac{2}{\alpha-2} \frac{\lfloor b_i \xi_i^{-4/\alpha}\rfloor^{-\alpha/2}}{b_i^{-\alpha/2} \xi^2_i} \right)
\nonumber \\
&=&
\frac{b_i\alpha}{\alpha-2}\xi_i^{2-4/\alpha} \cdot V( b_i \xi_i^{-4/\alpha})
\end{IEEEeqnarray}
where:
\begin{IEEEeqnarray}{rCl}\label{e: V define}
V(x)=\frac{\lfloor x\rfloor}{x}\cdot\left(1+\frac{2}{\alpha} \left( \frac{\lfloor x\rfloor^{-\alpha/2}}{x^{-\alpha/2}}-1\right)\right)
\end{IEEEeqnarray}
and we note in particular that:
\begin{IEEEeqnarray}{rCl}
\lim_{x\rightarrow \infty} V( x)=1.
\end{IEEEeqnarray}

Using the monotonic relation of the feedback rate ($F_i$) with $\xi_i^2$ we can prove (\ref{e: Th3 bound}) by inspecting the limit as $\xi_i$ approaches $0$. Combining (\ref{e: Ri from full version}), (\ref{e: F_i upper bound infinite network}) and (\ref{e: sum XMe second (ibtegral) bound}) and recalling that $\alpha>2$ we have:
\begin{IEEEeqnarray}{rCl}
\lim_{F_i\rightarrow \infty}\lim\limits_{\rho \rightarrow \infty}\frac{R_i}{\log_2F_i}&=&\lim_{\xi_i\rightarrow 0}\lim\limits_{\rho \rightarrow \infty}\frac{R_i}{\log_2F_i}
\nonumber \\
\TwoOneColumnAlternate{&&\hspace{-2cm}\ge}{&\ge&}
\lim_{\xi_i\rightarrow 0}\frac{E\left[\log_2\left(1+\frac{| \ell  _{i,i}|^2}{\frac{b_i\alpha}{\alpha-2}\xi^{2-4/\alpha}_i \cdot V( b_i \xi_i^{-4/\alpha})}\right)\right]}{\log_2\left(\frac{\alpha \log_2(e)}{2T}\left(  b_i \xi_i^{-4/\alpha}  -1\right)+\frac{ b_i \xi_i^{-4/\alpha}Q }{T}\right)}
\nonumber \\
\TwoOneColumnAlternate{&&\hspace{-2cm}=}{&=&}
\lim_{\xi_i\rightarrow 0}\frac{\log_2\left(\xi_i^{4/\alpha-2}\right)}{\log_2\left( \xi_i^{-4/\alpha}\right)}=\frac{\alpha}{2}-1.
\end{IEEEeqnarray}
which completes the proof of (\ref{e: Th3 bound}).

Finally, to prove (\ref{e: Th4 bound}), we again set the feedback rate to be proportional to the uplink rate, $F_i=r \cdot R_{\rm UL}$. Combining (\ref{e: Fi lower bound for fixed ratio}), which lower bounds $F_i$, with (\ref{e: F_i upper bound infinite network}), which upper bounds $F_i$, results in the following inequality on $\xi^2$:
\begin{IEEEeqnarray}{rCl}
\frac{\alpha \log_2(e)}{2T}\left(  b_i  \xi_i^{-4/\alpha}  -1\right)+\frac{ b_i  \xi_i^{-4/\alpha}Q }{T}\ge r\log_2 \rho  +g(\rho )\IEEEeqnarraynumspace
\end{IEEEeqnarray}
which can be stated as:
\begin{IEEEeqnarray}{rCl}\label{e: xi bound for infinite network}
\xi^2_i\le b_i^{\alpha/2} \left(\frac{2T r\log_2 \rho+2T g(\rho )+\alpha \log_2(e)  }{\alpha \log_2(e)    +2Q}  \right)^{-\alpha/2}.\IEEEeqnarraynumspace
\end{IEEEeqnarray}

To upper bound (\ref{e: V define}) we use:
\begin{IEEEeqnarray}{rCl}
V(x)\le1+\frac{2}{\alpha} \left( \frac{\lfloor x\rfloor^{-\alpha/2}}{x^{-\alpha/2}}-1\right)
\end{IEEEeqnarray}
which is upper bounded by:
\begin{IEEEeqnarray}{rCl}
V(x)\le c_0\triangleq 1+\frac{2}{\alpha} \left( 2^{\alpha/2}-1\right)
\end{IEEEeqnarray}
for any $x\ge 1$. Thus, using (\ref{e: sum XMe second (ibtegral) bound}) the sum of quantization error powers can be upper bounded by:
\begin{IEEEeqnarray}{rCl}\label{e: 2 another sum XMe bound}
\sum_{j=1}^{\infty}\XMe^2 _{i,j}&\le&
\frac{b_i\alpha}{\alpha-2}\xi^{2-4/\alpha}_i \cdot c_0
\end{IEEEeqnarray}
for any $\xi^2 \le b_i^{-\alpha/4}$.
Noting that the right hand side of  (\ref{e: 2 another sum XMe bound}) is monotonic increasing in $\xi^2_i$, we can substitute (\ref{e: xi bound for infinite network}) into (\ref{e: 2 another sum XMe bound}) and obtain:
\begin{IEEEeqnarray}{rCl}\label{e: Sum XMe final bound}
\TwoOneColumnAlternate{&&}{}\sum_{j=1}^{\infty}\XMe^2 _{i,j}
\TwoOneColumnAlternate{\nonumber \\&&\le}{&\le&} \frac{b_i\alpha  c_0}{\alpha-2}\bigg[ b_i^{\frac{\alpha}{2}} \bigg(\frac{2T r\log_2 \rho+2T g(\rho )+\alpha \log_2(e)  }{\alpha \log_2(e)    +2Q}  \bigg)^{-\frac{\alpha}{2}}\bigg]^{1-\frac{2}{\alpha}}
\nonumber \\
\TwoOneColumnAlternate{&&=}{&=&}
\frac{\alpha c_0}{\alpha-2}
b_i^{\frac{\alpha}{2}} \left(\frac{2T r\log_2 \rho+2T g(\rho )+\alpha \log_2(e)  }{\alpha \log_2(e)    +2Q}  \right)^{1-\frac{\alpha}{2}}.
\end{IEEEeqnarray}
Noting that $\lim_{N \rightarrow \infty}\lim_{\rho \rightarrow\infty} g(\rho)$ is finite in any physical channel model we have:
\begin{IEEEeqnarray}{rCl}\label{e: Sum XMe ratio bound}
\lim_{\rho\rightarrow\infty}\frac{\sum_{j=1}^{\infty}\XMe^2 _{i,j}}{(\log_2\rho)^{1-\frac{\alpha}{2}}}\le
\frac{\alpha c_0}{\alpha-2}
b_i^{\frac{\alpha}{2}} \left(\frac{2T r }{\alpha \log_2(e)    +2Q}  \right)^{1-\frac{\alpha}{2}}.\IEEEeqnarraynumspace
\end{IEEEeqnarray}

To conclude, we divide the rate lower bound, (\ref{e: Ri from full version}),  by $\log_2\log_2 \rho $ and take the limit as $\rho $ goes to infinity. Recalling that $\alpha>2$ and substituting (\ref{e: Sum XMe ratio bound}) results in (\ref{e: Th4 bound}) and completes the proof of Theorem \ref{Th: Fixed feedback rate infinite}.\hfill$\blacksquare$

%%%%%%%%%%%%%%%%%%%%%%%%%%%%%%%%%%%%
\section{Numerical example}\label{sec: numeric}
In this section we illustrate the results derived above, by presenting simulation results for a system that performs interference mitigation with variable feedback rate. The emphasis in this section is on proving that the rate tradeoff is indeed achievable in practical systems. Therefore, we simulate a simple system that applies very basic  methods for quantization and for interference cancellation. The system is based on the lattice precoding approach described in \cite{Bergel2014SPAWC}. The results show that even this simplified system, although suboptimal, yields the results expected from our analysis, and in particular the rate tradeoff ratio of (\ref{e: inifinite network slope apx}).

\TwoOneColumnAlternate{
\FigDat{System}{0.55}
}{
\FigDat{System}{0.8}
}

\subsection{Simulated system}
\subsubsection{Feedback scheme}
We adopt the suboptimal scalar quantization scheme of Sub-subsection \ref{subsub: scalar quantization}. The quantization is based on a fixed step size of $\Delta$ (for the real and imaginary parts of the channel, separately). Thus the quantization noise is $\xi^2=\Delta^2/6$, and we decide to send feedback to a BS if its received power, $\sigma^2_{i,j}$, is larger than $\xi^2$. As stated above, before quantization we add a dither signal which is uniformly distributed in the range $[-\Delta/2,\Delta/2)$. This dither is known to the BS and is removed from the quantized values. The quantization values, from the real and imaginary parts of the channel gain from each participating BS, are compressed using  variable rate Huffman coding \cite{huffman1952method} (resulting in an average feedback rate which is close to the entropy of the quantized values).

\subsubsection{Transmission scheme}

Again we turn to a one dimensional simplification, and adopt a suboptimal scalar modulo scheme. To generate the data vector, $\xv$, we start with source symbols $\vv$ that are generated with real and imaginary parts, each distributed uniformly over the range $\mathcal{V}=\big[-\sqrt{3/2},\sqrt{3/2}\, \big)$. We also generate a dithering signal $\uv$ with iid elements and the same uniform distribution.
The transmitted symbols are generated by:
\begin{IEEEeqnarray}{rCl}
\xv=\Qm^H \sv
\end{IEEEeqnarray}
where the matrix $\Qm$ comes from the LQ decomposition of the quantized channel matrix (see (\ref{e: LQ decomp})), and the $i$-th element of the vector $\sv$ is calculated by:
\begin{IEEEeqnarray}{rCl}
s_i=\left[v_i-\sum_{j< i}  a_i \ell_{i,j} s_j  -u_i\right] \mod \mathcal{V}
\end{IEEEeqnarray}
where 
\begin{IEEEeqnarray}{rCl}
a_i =\frac{\ell_{i,i}^*}{\rho|\ell_{i,i}|^2+\rho\sum_j \min\{\xi^2,\sigma^2_{i,j}\}+1}
\end{IEEEeqnarray}
and the modulo operation is defined for complex signals as:
\begin{IEEEeqnarray}{rCl}
x \mod \mathcal{V}\triangleq\left(\Re( x) \mod \mathcal{V} \right)+j\cdot\left(\Im( x) \mod \mathcal{V} \right).\IEEEeqnarraynumspace
\end{IEEEeqnarray}
and for real signals as:
\begin{IEEEeqnarray}{rCl}
x \mod \mathcal{V}\triangleq \left( x + \sqrt{\frac{3}{2}} \mod \sqrt{6} \right)-\sqrt{\frac{3}{2}}.
\end{IEEEeqnarray}

\subsubsection{Reception scheme}
The receiver performs the preprocessing:
\begin{IEEEeqnarray}{rCl}
y_i^\prime=[a_i y_i+u_i] \mod \mathcal{V}
\end{IEEEeqnarray}
The achievable rate is evaluated through empiric calculation of the mutual information between $y_i^\prime$ and $v_i$.

\TwoOneColumnAlternate{
\FigDat{mult_BS_select}{0.5}
}{
\FigDat{mult_BS_select}{0.8}
}
\subsection{Simulation results}
We first simulated a network with $55$ BS placed over a hexagonal grid as depicted in Fig. \ref{f:system model}, where each BS is equipped with a single antenna. The center mobile is located at the border point of $3$ cells. The path loss exponent is set to $\alpha=4$. To simplify the simulation we evaluate the channel powers for the middle mobile ($\sigma^2_{i,j}$ for $i=1$) and these powers are used in a cyclic manner for all mobiles. The transmitted signal is set so that the ratio between the signal power and the noise power is $30$dB (so that the simulation is interference limited). The interference from the unsimulated (infinite number of) BSs is simulated by an additional Gaussian noise with a power $0.027\cdot \rho$ (which was numerically calculated based on a huge grid). Thus, the maximal SINR achievable in this simulation is $16$dB.

Fig. \ref{f:multch fig} depicts the average spectral efficiency per user at the downlink as a function of the amount of spectral efficiency per user at the uplink that is devoted to feedback. Following the ETSI Pedestrian B channel example presented above \cite{etsi1998101}, the feedback spectral efficiency is evaluated for $T=180$ symbols. The figure presents the results for different levels of BS cooperation. With no BS cooperation, the feedback is not required, and hence the achievable spectral efficiency is constant for all feedback rates. As all mobiles in the simulation are located at cell edges, their average spectral efficiency is quite low ($0.56$ Bps/Hz).

\TwoOneColumnAlternate{
\FigDat{Performance}{0.5}
}{
\FigDat{Performance}{0.8}
}

In the simulated setup, each mobile is located at the same distance from its $3$ nearest BSs. Thus, we first consider cooperation of $3$ BSs. Such level of cooperation is shown to be very useful for low feedback rates.
Fig. \ref{f:multch fig} shows that the downlink rate increases almost linearly with the feedback rate, up to a rate which is almost $3$ times that of the no-cooperation rate. But, as the feedback rate increases, the interference from other transmitters becomes dominant, and the rate cannot increase above $1.5$ Bps/Hz.

\TwoOneColumnAlternate{
\FigDat{VsRho}{0.5}
}{
\FigDat{VsRho}{0.8}
}

Higher rates are achieved using $6$ BSs cooperation, but only when the feedback rate is high enough. For feedback rates below $0.12$ Bps/Hz, the figure shows that sending feedback to $6$ BSs is not optimal and the feedback to the further $3$ BSs is `wasted'. The same results hold also for higher level of cooperation: We see that $12$ BSs cooperation is optimal for feedback rates higher than $0.25$ Bps/Hz, and $21$ BSs cooperation is optimal for feedback rates higher than $0.5$ Bps/Hz.

The optimal performance of the simulated systems is given by the convex hull of all the curves in Fig. \ref{f:multch fig}. However, it is difficult to predict in advance what is the optimal number of BSs to feedback for a given feedback rate. Thus, we turn to the suboptimal solution of Subsection \ref{subsub: proof inifinite}, and choose the number of BSs as if the quantization is optimal. The resulting spectral efficiency is depicted in Fig. \ref{f:balance fig}. 
 As expected, the exchange ratio of uplink to downlink rate is quite favorable.  For example, the right most point in Fig. \ref{f:balance fig} shows that spending a spectral efficiency of $0.7$ Bps/Hz of the uplink for feedback, improves the downlink spectral efficiency by a factor of $6$, from $0.56$ Bps/Hz to $3.5$ Bps/Hz. Thus each bit of feedback on the uplink increases the downlink by more than $4$ bits.

The simulation results would be expected to converge to the behavior predicted by (\ref{e: inifinite network slope apx}) for a sufficiently large feedback rate. Evaluating the curve slope using (\ref{e: inifinite network slope apx}) for the cases of $L_i(F_i)=6, 12, 21$ (using $Q=2$) results with $8.9$, $4.4$ and $2.5$, respectively. For convenience Fig. \ref{f:balance fig} shows linear lines with these slopes at the relevant points on the performance curve. As expected, for $6$ BSs the approximation is not very good. But, as the number of BSs that receive feedback increases, the slope approximation becomes very accurate.

The figure also shows the performance of the same network when each BS is equipped with $2$ antennas. As expected, the spectral efficiency of each user is higher than with single antenna BSs. In this case, the convergence to the slope approximation of (\ref{e: inifinite network slope apx}) is slower, but one can see that for $12$ BSs ($24$ antennas) the approximation is already quite accurate. Also, one should note that even while sending feedback to $24$ antennas, the exchange ratio of uplink to downlink rate is still favorable. 

For reference, the figure also shows the performance of the linear ZF scheme. As stated above, the gap between the DP and the ZF scheme increases as the number of cooperating BS increases. Hence, the performance of the ZF scheme is quite disappointing for high feedback rates.

To further illustrate our results, Fig. \ref{f:VsRho} depicts the average downlink spectral efficiency as a function of the ratio between the power of the signal from the nearest BS and the noise power (SNR). The results are depicted for a network in which the channel gains are identical to those presented above, but only for the $6$ BSs which are the nearest to each mobile. In this setup, all further BSs have no effect on the mobile. Thus, this network matches the conditions of Theorem \ref{Th: Fixed feedback rate finite}, with $L_i=6$. The solid lines depict the achievable performance for feedback with fixed rates. As predicted by the first part of Theorem \ref{Th: Fixed feedback rate finite} (Equation (\ref{e: th1_bound})), for high SNR the differences between the different curves is nearly constant. The $x$-marks depict the performance when the feedback rate takes $3\%$ of the uplink rate (where we set $\kappa_{\rm UL}$ of Equation (\ref{e:k_UL}) to be $0.1$). The dashed line shows the slope predicted by the second part of Theorem \ref{Th: Fixed feedback rate finite} (Equation (\ref{e: Th2 bound})). As can be seen, the predictions of Theorem \ref{Th: Fixed feedback rate finite} are very accurate at the high SNR regime.
%%%%%%%%%%%%%%%%%%%%%%%%%%%%%%%%%%%%
\section{conclusions}\label{sec: discussion}
In this work we address the relation between the downlink rate and the uplink rate in an FDD cellular network, and quantify the tradeoff between these two rates.
The presented results demonstrate that a flexible control of the feedback rate can result in an efficient mechanism for uplink-downlink rate balancing. Previous research has already established the general understanding that increasing the feedback rate (over the uplink) will increase the downlink rate. In this work we have quantified this tradeoff and demonstrated the capability to dynamically balance the uplink and the downlink rates. We also extended the analysis to infinite networks and showed that the overall network downlink rate can grow unbounded with SNR, by using a constant fraction of the uplink for feedback.

The presented lower bounds showed that in the interference limited regime, there is an exchange ratio between uplink rate and downlink rate.
The rate tradeoff ratio is constant in finite
networks and equals to the ratio between the coherence block length, $T$, and the total number of BS antennas in the network. In infinite networks this ratio is a decreasing function of the rate. Yet we showed that the rate tradeoff ratio in an infinite network can be approximated by the ratio between the coherence block length and the number of antennas in the feedback set, multiplied by a constant. As the coherence time in cellular systems is typically quite large, while the number of cooperating BSs is more limited, the exchange ratio of uplink to downlink rate is quite favorable.

The ability to trade uplink rate for downlink rate with a favorable ratio is very encouraging. This ability can allow cellular operators to balance the uplink and downlink rate in a flexible manner, and hence significantly improve the network efficiency.

The results presented in this work raise several issues that deserve further study. One such issue is the capability of each mobile to accurately estimate the channel from many antennas. When the number of estimated  channels becomes large, the estimation capability is limited, which will affect the network capacity. Furthermore, even if the mobile manages the channel estimation, the feedback must reach the BSs before the channel changes. For fast moving mobiles this may not be feasible, and the feedback may be outdated by the time that it is used. Future work can address this case 
by incorporating interference alignment approaches of the kind suggested by Maddah-Ali and Tse \cite{maddah2012completely}.

Another issue is the user scheduling method. While known methods suggest to choose mobiles based on (accurate or coarse) CSI, this work raises an additional criterion: mobiles should be selected in a way that reduces the required feedback rate (i.e., the sum of the feedback rate from the mobile for all quantized channels). Note that these rates change quite slowly, and hence can be assumed to be known to the BSs. Thus, the use of this criterion can significantly reduce the required feedback rate for a given downlink throughput.

\section*{Acknowledgment}
We wish to thank the AE and the reviewers whose comments enhanced the technical quality of the contribution. We also wish to thank Dr. Shimon Moshavi for his valuable advice and comments that helped us improve this manuscript. 

%%%%%%%%%%%%%%%%%%%%%%%%%%%%%%%%%%%%%%%%%%%%
\begin{appendices}

%%%%%%%%%%%%%%%%%%%%%%%%%%%%%%%%%%%%
\section{Proof of Proposition \ref{cor: cor 1}}\label{app: proof of corrolary on rate slope}
This proof reuses much of the proof of Theorem \ref{Th: Fixed feedback rate infinite}, and the considered transmission scheme is the scheme of Theorem \ref{Th: Fixed feedback rate infinite}. A closed form expression of the suggested bound, $\tilde R_i(\rho)$ is difficult to obtain. Instead, we use again the approach of Theorem \ref{Th: Fixed feedback rate infinite} and study the performance as a function of the target quantization error $\xi_i^2$.  As stated above, the quantization error is monotonic decreasing with the SNR, and hence the asymptotic behavior for $\rho\rightarrow \infty$ is identical to the behavior for $\xi_i^2\rightarrow0$.

We use:
\begin{IEEEeqnarray}{rCl}\label{e: F approx}
 \tilde F_i \triangleq \frac{b_i \xi_i^{-4/\alpha}}{2T} \left(\alpha \log_2(e)+2Q\right) -\frac{\alpha \log_2(e)}{2T}\IEEEeqnarraynumspace
\end{IEEEeqnarray}
recalling that  $F_i\le \tilde F_i$ was proved in (\ref{e: F_i upper bound infinite network}). Substituting (\ref{e: sum XMe second (ibtegral) bound}) into (\ref{e: Ri from full version}) we have for $ b_i \xi_i^{-4/\alpha}\ge 1$:  
\begin{IEEEeqnarray}{rCl}
 R_i \ge E\left[\log_2\left(\frac{\rho | \ell  _{i,i}|^2}{1+\rho \cdot\frac{b_i\alpha}{\alpha-2}\xi_i^{2-4/\alpha} \cdot V( b_i \xi_i^{-4/\alpha})}\right)\right]
\end{IEEEeqnarray}
where $V()$ is given in (\ref{e: V define}). Noting that $V(x)$ is monotonic decreasing in ${\lfloor x\rfloor}/{x}$, we can upper bound $V(x)$ by:
\begin{IEEEeqnarray}{rCl}
V(x)<\frac{x -1}{x}\cdot\left(1+\frac{2}{\alpha} \left( \frac{(x-1)^{-\alpha/2}}{x^{-\alpha/2}}-1\right)\right).
\end{IEEEeqnarray}
Thus, we can satisfy $R_i\ge \tilde R_i$ by defining 
\begin{IEEEeqnarray}{rCl}\label{e: R approx}
 \tilde R_i = E\left[\log_2\left(\frac{\rho | \ell  _{i,i}|^2}{1+\rho \cdot\frac{\alpha}{\alpha-2}\cdot \tilde G(\xi_i^2)}\right)\right]
\end{IEEEeqnarray}
and
\begin{IEEEeqnarray}{rCl}\label{e: G define}
\tilde G(\xi_i^2)&\triangleq& b_i\xi_i^{2-4/\alpha} \cdot \tilde V( b_i \xi_i^{-4/\alpha})
\TwoOneColumnAlternate{\\ \nonumber &&\hspace{-1.3cm}=}{\nonumber \\ &=&}
 \xi_i^{2}(b_i \xi_i^{-4/\alpha} -1)\cdot\left(1+\frac{2}{\alpha} \left( \frac{(b_i \xi_i^{-4/\alpha}-1)^{-\alpha/2}}{b_i^{-\alpha/2} \xi_i^{2}}-1\right)\right).
\end{IEEEeqnarray}
Note that the derivation of both (\ref{e: F approx}) and (\ref{e: R approx}) inherently assumes  (see  (\ref{e: Li upper bound})):
\begin{IEEEeqnarray}{rCl}\label{e: L approx}
 \tilde L_i \triangleq \lfloor b_i \xi_i^{-4/\alpha} \rfloor .
\end{IEEEeqnarray}

Taking the derivative of $\tilde R_i$ with respect to $\tilde F_i$ we have:
\begin{IEEEeqnarray}{rCl}
\frac{d\tilde R_i}{d\tilde F_i}&=&\frac{{d\tilde R_i}/{d(\xi_i^2)}}{{d\tilde F_i}/{d(\xi_i^2)}}
\TwoOneColumnAlternate{\\ \nonumber &&\hspace{-.8cm}=}{\nonumber \\ &=&}
\frac{E\left[\frac{1}{\rho} \frac{d \rho}{d(\xi_i^2)}-\frac{\frac{\alpha}{\alpha-2}}{1+\rho\frac{\alpha}{\alpha-2} \tilde G(\xi_i^2)}\left(\frac{d\rho}{d(\xi_i^2)}  \tilde G(\xi_i^2)+\rho\frac{d   \tilde G(\xi_i^2)}{d(\xi_i^2)}\right)\right]}{-\frac{2}{\alpha}\frac{b_i \xi_i^{-4/\alpha-2}}{2T} \left(\alpha \log_2(e)+2Q\right) \cdot\frac{1}{\log_2\left(e \right)}}.
\end{IEEEeqnarray}
From (\ref{e: Sum XMe ratio bound}) we have $\rho \frac{\alpha}{\alpha-2} \tilde G(\xi_i^2)=\rho \cdot\sum_{j=1}^M \XMe_{i,j}\rightarrow\infty$ as $\xi_i^2\rightarrow 0$. Thus,
\begin{IEEEeqnarray}{rCl}\label{e: inifinite network slope step}
\TCee\lim_{\xi_i^2\rightarrow0}\frac{  \tilde L_i(\rho) }{T }\frac{d\tilde R_i(\rho)}{d \tilde F_i}
\TwoOneColumnAlternate{\nonumber \\&&=}{&=&}
\lim_{\xi_i^2\rightarrow0}\frac{  \tilde L_i(\rho) }{T }\frac{E\left[\frac{1}{\rho} \frac{d \rho}{d(\xi_i^2)}-\frac{1}{\rho \tilde G(\xi_i^2)}\left(\frac{d\rho}{d(\xi_i^2)}  \tilde G(\xi_i^2)+\rho\frac{d   \tilde G(\xi_i^2)}{d(\xi_i^2)}\right)\right]}{-\frac{2}{\alpha}\frac{b_i \xi_i^{-4/\alpha-2}}{2T} \left(\alpha \log_2(e)+2Q\right)\cdot\frac{1}{\log_2\left(e \right)} }
\nonumber \\
\TwoOneColumnAlternate{&&=}{&=&}
\lim_{\xi_i^2\rightarrow0}\frac{  \tilde L_i(\rho) }{T }\frac{E\left[\frac{1}{ \tilde G(\xi_i^2)}\frac{d   \tilde G(\xi_i^2)}{d(\xi_i^2)}\right]\log_2\left(e \right)}{\frac{2}{\alpha}\frac{b_i \xi_i^{-4/\alpha-2}}{2T} \left(\alpha \log_2(e)+2Q\right) }.
\end{IEEEeqnarray}
Using (\ref{e: G define}) and some algebra gives the derivative:
\begin{IEEEeqnarray}{rCl}\label{e: G derivative}
\frac{1}{ \tilde G(\xi_i^2)}\frac{d   \tilde G(\xi_i^2)}{d(\xi_i^2)}
&=&\frac{(1-\frac{2}{\alpha})b_i \xi_i^{-4/\alpha} -1}{\xi_i^{2}(b_i \xi_i^{-4/\alpha} -1)}
\nonumber \\&&
\TwoColHspace{-1cm}+\frac{(b_i \xi_i^{-4/\alpha}-1)^{-\alpha/2-1}}{(1+\frac{2}{\alpha})b_i^{-\alpha/2} \xi_i^{4}+\frac{2}{\alpha} \xi_i^2(b_i \xi_i^{-4/\alpha}-1)^{-\alpha/2}}
\IEEEeqnarraynumspace
\end{IEEEeqnarray}
and we note that
\begin{IEEEeqnarray}{rCl}
\lim_{\xi_i^2\rightarrow 0}\frac{\xi_i^{2}}{ \tilde G(\xi_i^2)}\frac{d   \tilde G(\xi_i^2)}{d(\xi_i^2)}=1-\frac{2}{\alpha}.
\end{IEEEeqnarray}
Substituting (\ref{e: G derivative}) and (\ref{e: L approx}) into (\ref{e: inifinite network slope step}) and evaluating the limit leads to (\ref{e: inifinite network slope}) and completes the proof of the proposition. \hfill$\blacksquare$
\end{appendices}

\bibliographystyle{IEEEtran}
\bibliography{../../../Bergel_all_bib}
%\bibliography{ULDL_Jul_18}
\clearpage

%\TwoOneColumnAlternate{}{\FigDat{System}{0.9}\FigDat{Performance}{0.9}}

\end{document}